\documentclass[apj]{emulateapj}

\usepackage{graphicx,amssymb}

\newcommand{\kms}{\ifmmode {\rm km\,s}^{-1} \else km\,s$^{-1}$\fi}

\newcommand{\ltapprox}{\scriptsize\hbox{\raise0.5ex\hbox{$<$}\kern-0.75em
\lower0.5ex \hbox{$\sim$}}\normalsize}

\slugcomment{Draft version 1.5 [\today]} 
\shorttitle{Excited-state OH masers} 
\shortauthors{Pihlstr\"om et al.}

\begin{document}


\title{Excited-state OH Masers and Supernova Remnants}

\author{Ylva M.\ Pihlstr\"om}\affil{Department of Physics and
  Astronomy, MSC07 4220, University of New Mexico, Albuquerque, NM
  87131}

\author{Vincent L.\ Fish\altaffilmark{1,2} \and Lor\'ant O.\
  Sjouwerman}\affil{National Radio Astronomy Observatory, PO Box 0,
  Socorro, NM 87801}

\author{Laura K.\ Zschaechner}\affil{Department of Physics and
  Astronomy, MSC07 4220, University of New Mexico, Albuquerque, NM
  87131}

\author{Philip B.\ Lockett}\affil{Centre College, 600 West Walnut
  Street, Danville, KY 40422}

\author{Moshe Elitzur}\affil{Department of Physics and Astronomy,
  University of Kentucky, Lexington, KY 40506-0055}

\altaffiltext{1}{Jansky Fellow}
\altaffiltext{2}{Current address: Massachusetts Institute of
  Technology, Haystack Observatory, Route 40, Westford, MA 01886}


\begin{abstract}

  The collisionally pumped, ground-state 1720\,MHz maser line of OH is
  widely recognized as a tracer for shocked regions and observed in
  star forming regions and supernova remnants. Whereas some lines of
  excited states of OH have been detected and studied in star forming
  regions, the subject of excited-state OH in supernova remnants --
  where high collision rates are to be expected -- is only recently
  being addressed. Modeling of collisional excitation of OH
  demonstrates that 1720, 4765 and 6049\,MHz masers can occur under
  similar conditions in regions of shocked gas. In particular, the
  6049 and 4765\,MHz masers become more significant at increased OH
  column densities where the 1720\,MHz masers begin to be quenched. In
  supernova remnants, the detection of excited-state OH line maser
  emission could therefore serve as a probe of regions of higher
  column densities. Using the Very Large Array, we searched for
  excited-state OH in the 4.7, 7.8, 8.2 and 23.8\,GHz lines in four
  well studied supernova remnants with strong 1720\,MHz maser emission
  (Sgr\,A\,East, W\,28, W\,44 and IC\,443). No detections were made,
  at typical detection limits of around 10 mJy\,beam$^{-1}$. The
  search for the 6\,GHz lines were done using Effelsberg since the VLA
  receivers did not cover those frequencies, and are reported on in an
  accompanying letter \citep{fish07}. We also cross-correlated the
  positions of known supernova remnants with the positions of
  1612\,MHz maser emission obtained from blind surveys. No probable
  associations were found, perhaps except in the Sgr\,A\,East
  region. The lack of detections of excited-state OH indicates that
  the OH column densities suffice for 1720\,MHz inversion but not for
  inversion of excited-state transitions, consistent with the expected
  results for C-type shocks.

\end{abstract}


\maketitle


\section{Introduction}\label{intro}
Masers observed in the 1720\,MHz satellite line of OH are often
associated with supernova remnants (SNRs). They originate in the
shocked region where the expanding SNR collides with a molecular
cloud. During the collision, a non-dissociative C-type shock can
produce the temperature and density conditions required for 1720\,MHz
maser emission to occur \citep{wardle99, lockett99}. The C-type
shock model requires shock speeds of the order of 25 km\,s$^{-1}$, and
shock chemistry predicts the conditions to be right for masers behind
the shock wave. Such velocities and spatial positions are in good
agreement with previous observations of SNRs
\citep[e.g.,][]{claussen97, yusef-zadeh03b, frail98}. A natural, next
step would be to perform very long baseline interferometric (VLBI)
observations to measure the physical sizes of the masing regions, and
to track the SNR expansion with time. However, a major hurdle at
1720\,MHz is the angular broadening due to interstellar
scattering. For example, in the direction of the Galactic center,
pronounced scattering ($\Theta_{obs}\sim 500$ mas at 1.6\,GHz) has
been measured \citep{vanlangevelde92}. To overcome this problem,
\citet{hoffman03} conducted MERLIN and VLBA observations of the
1720\,MHz masers in IC\,443, positioned at a Galactic longitude
believed to be little affected by scattering. In another project,
\citet{claussen02} used a nearby pulsar to estimate the interstellar
scattering effect close to W\,28.

A more general method that could work at any position on the sky would
be using higher frequency transitions tracing the same type of gas
(shocked by the SNR/cloud collisions). Higher frequency masers are
less susceptible to scattering effects, which scale as
$\lambda^{2}$. Finding higher frequency lines would be particularly
interesting for the Sgr\,A\,East region, since some 1720\,MHz masers
found here have velocities very offset from what is expected in the
SNR model \citep{yusef-zadeh96, karlsson03}. These offset velocity
masers coincide on the sky with the circumnuclear disk and might
therefore arise under different conditions. Proper motion studies of
masers in the Sgr\,A complex would thus indicate whether the
kinematics really differ for the offset velocity masers and the
Sgr\,A\,East masers. Higher frequency excited-state OH masers thus
could be used for astrometric and proper motion VLBI studies along the
heavily scattered Galactic plane and in the Galactic center.

Modeling of the three lower rotational states of OH has demonstrated
that in star forming regions (SFRs) satellite line 1720 and 4765\,MHz,
and mainline 6035\,MHz masers can occur under similar conditions in
regions of shocked gas \citep{gray91, gray92}. The 4765 and 6035\,MHz
masers become more significant at increased column densities where the
1720\,MHz masers begin to switch off. Detection of excited-state OH
maser emission could therefore serve as a probe of regions of higher
OH column densities in SNRs ($N_{\rm OH}\sim 3\times
10^{17}$\,cm$^{-2}$ instead of $N_{\rm OH}\sim 3\times
10^{16}$\,cm$^{-2}$ for the 1720\,MHz masers). However, calculations
of the $N_{\rm OH}$ resulting from C-type shocks are only able to
produce relatively low values, $N_{\rm OH} \simeq 10^{16}$\,cm$^{-2}$
\citep{lockett99, wardle99} which may be a limiting factor in the
formation of excited-state OH masers in SNRs. It should be noted that
detailed observations of OH and H$_2$O in IC\,443 require an OH
production scenario involving both J- and C-type shocks
\citep{snell05, hewitt06}, indicating that estimating the OH column
density could be more complicated than previously thought.

In this work we present the details and results of our search for
higher frequency OH transitions in SNRs with the Very Large Array
(VLA). We also report on the results of a literature search for
collisionally excited 1612\,MHz masers associated with SNRs.

\begin{deluxetable*}{llllrccr}[t]
\tabletypesize{\footnotesize}
\tablecolumns{8}
\tablewidth{0pt}
\tablecaption{Observational parameters}
\tablehead{\colhead{Pointing}                     &
  \multicolumn{2}{c}{RA   (J2000) Dec (J2000)}   &
  \colhead{Transition}                          &
  \colhead{V$_{\rm sys}$\tablenotemark{a}}                        &
  \colhead{$\Delta$V$_{\rm ch}$\,\tablenotemark{b}}        &
  \colhead{Beam size\tablenotemark{c}}           &
  \colhead{Mean rms\tablenotemark{c}}           \\
  \colhead{}                                    &
  \colhead{{h}\phn{m}\phn{s}}                   &
  \colhead{\phn{\arcdeg}~\phn{\arcmin}~\phn{\arcsec}} &
  \colhead{MHz}                                &
  \colhead{km~s$^{-1}$}                         &
  \colhead{km~s$^{-1}$}                         &
  \colhead{\phn{\arcsec}$\times$\phn{\arcsec}}  &
  \colhead{mJy beam$^{-1}$}                      }

\startdata
IC443 & 06 16 43.61 & $+$22 32 39.78 & 4750.7, 4660.2, 4765.6&$-$4.6&0.9 & 5.1$\times$4.3 & 13.2 \\
 &    &  & 7761.7, 7820.1, 7832.0, 7749.9  & $-$4.6 & 0.6 & 3.0$\times$2.8 & 8.4 \\
 &    &  & 8135.9, 8189,6, 8207.4, 8118.1  & $-$4.6 & 0.5 & 3.0$\times$2.7 & 6.5 \\
Sgr\,A\,East & 17 45 44.31 & $-$29 01 18.34 & 4750.7 & 64.0 & 0.8 & 7.7$\times$3.4 & 13.8 \\
 &              &                 & 4750.7 & 132.0 & 0.8 & 7.7$\times$3.4 & 14.5 \\
 &    &                   & 4660.2, 4765.6 & 64.0 & 0.8 & 7.7$\times$3.4 & 14.1 \\
 &    &  & 7761.7, 7820.1, 7832.0, 7749.9 &  64.0 & 0.5 & 7.0$\times$2.5 & 11.4 \\
 &    &  & 8135.9, 8189,6, 8207.4, 8118.1 &  64.0 & 0.5 & 5.7$\times$2.6 & 9.5 \\
 & & & 23817.9, 23826.6, 23838.9, 23805.3 &  64.0 & 0.7 & 1.4$\times$0.8 & 9.4 \\
Sgr\,A\,East & 17 45 40.62 & $-$28 59 43.98 & 4750.7 & 64.0 & 0.8 & 5.8$\times$3.9 & 15.1 \\
 &              &                 & 4750.7 & 132.0 & 0.8 & 5.8$\times$3.9 & 14.1 \\
 &    &                   & 4660.2, 4765.6 & 132.0 & 0.8 & 5.8$\times$3.9 & 14.6 \\
 &    &  & 7761.7, 7820.1, 7832.0, 7749.9  & 132.0 & 0.5 & 5.4$\times$2.6 & 11.9 \\
 &    &  & 8135.9, 8189,6, 8207.4, 8118.1  & 132.0 & 0.5 & 5.3$\times$2.5 & 9.5 \\
 & & & 23817.9, 23826.6, 23838.9, 23805.3 & 132.0 & 0.7 & 1.2$\times$0.9 & 9.5 \\
W28A &18 00 45.55&$-$23 17 43.33& 4750.7 & 6.3  & 0.9 & 5.9$\times$4.0 & 8.6 \\
&              &                & 4750.7 & 76.3 & 0.9 & 5.9$\times$4.0 & 8.8 \\
&              &        & 4660.2, 4765.6 &  6.3 & 0.9 & 6.2$\times$4.2 & 8.7 \\
&    &  & 7761.7, 7820.1, 7832.0, 7749.9 &  6.3 & 0.6 & 4.2$\times$2.6 & 9.1 \\
&    &  & 8135.9, 8189,6, 8207.4, 8118.1 &  6.3 & 0.6 & 4.5$\times$2.6 & 7.4 \\
W28CD&18 01 39.35&$-$23 25 01.97 & 4750.7 & 14.1 & 0.9 & 5.7$\times$4.0 & 9.2 \\
&              &                 & 4750.7 & 84.1 & 0.9 & 5.9$\times$4.1 & 8.0 \\
&              &         & 4660.2, 4765.6 & 14.1 & 0.9 & 6.2$\times$3.9 & 8.6 \\
&     &  & 7761.7, 7820.1, 7832.0, 7749.9 & 14.1 & 0.6 & 4.8$\times$2.6 & 9.1 \\
&     &  & 8135.9, 8189,6, 8207.4, 8118.1 & 14.1 & 0.5 & 4.1$\times$2.8 & 7.2 \\
W28EF&18 01 51.64&$-$23 18 21.02 & 4750.7 & 11.7 & 0.9 & 6.7$\times$3.6 & 8.3 \\
&              &                 & 4750.7 & 81.7 & 0.9 & 6.9$\times$3.5 & 9.2 \\
&     &                  & 4660.2, 4765.6 & 11.7 & 0.9 & 7.2$\times$4.3 & 8.9 \\
&     &  & 7761.7, 7820.1, 7832.0, 7749.9 & 11.7 & 0.6 & 4.6$\times$2.8 & 9.1 \\
&     &  & 8135.9, 8189,6, 8207.4, 8118.1 & 11.7 & 0.5 & 5.0$\times$2.6 & 7.1 \\
W44EF\tablenotemark{d} & 18 56 33.02 & $+$01 27 54.40 & 4750.7 & 6.3  & 0.9 & 5.3$\times$3.9 & 8.9 \\
 &              &                 & 4750.7 & 76.3 & 0.9 & 5.3$\times$3.9 & 10.3 \\
 &    &                   & 4660.2, 4765.6 &  6.3 & 0.9 & 6.2$\times$3.7 & 8.9 \\
W44E\tablenotemark{d}  & 18 56 29.14 & +01 29 14.15 & 7761.7, 7820.1, 7832.0, 7749.9 & 6.3 & 0.6 &3.9$\times$2.5&8.1 \\
 &    &  & 8135.9, 8189,6, 8207.4, 8118.1 &  6.3 & 0.6 & 3.9$\times$2.4 & 6.3 \\
W44F\tablenotemark{d} & 18 56 36.90 & $+$01 26 34.65 & 7761.7, 7820.1, 7832.0, 7749.9 & 6.3 & 0.6 &4.4$\times$2.4&8.0 \\
&    &  & 8135.9, 8189,6, 8207.4, 8118.1 &  6.3 & 0.6 & 4.1$\times$2.4 & 6.5 \\
 \enddata

 \tablenotetext{a}{The observed bandwidth was centered on the systemic
   LSR velocity, and covered $\sim$96, 58, 55 and 36 km\,s$^{-1}$ for
   the 4.7, 7.8, 8.2 and 23.8\,GHz lines respectively. }
 \tablenotetext{b}{Velocity resolution, with no on-line Hanning
   smoothing applied during the observations} \tablenotetext{c}{The
   transitions were imaged separately, but lines close in frequency
   have similar beam sizes and channel rms. The listed value is the
   mean of the transitions.}  \tablenotetext{d}{At lower frequencies
   W44E and W44F were observed in one beam centered at their mean
   position, W44EF. At higher frequencies, they were observed
   separately as W44E and W44F.}
\end{deluxetable*}

\section{Data collection}

\subsection{Observations of  excited-state OH lines}\label{observations}
\begin{figure}[t]
\resizebox{9cm}{!}{\includegraphics{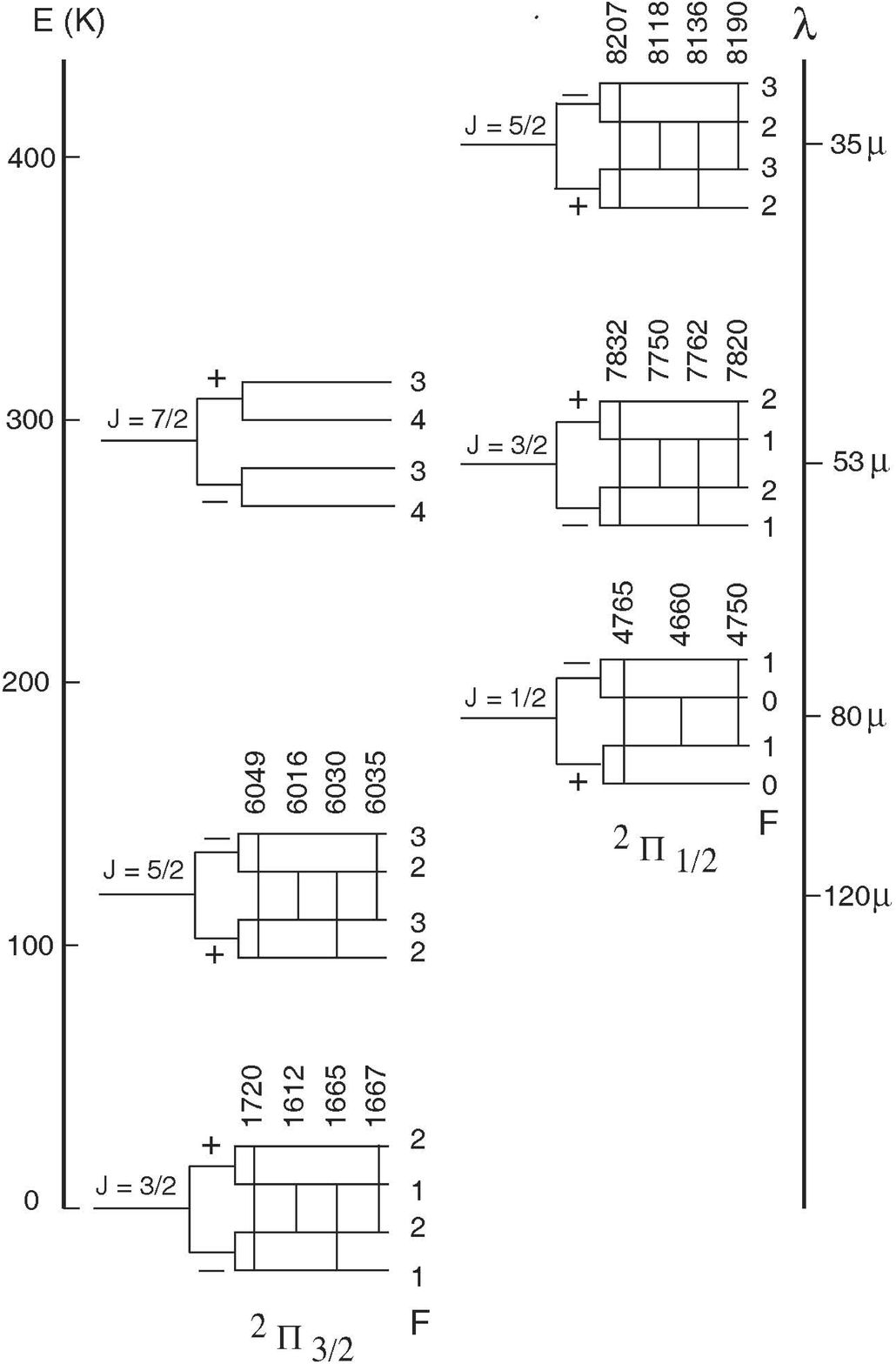}}
\caption{Energy level diagram for the OH molecule, showing transitions
  presently observable with the (E)VLA with excitation levels less
  than 500\,K. Transition frequencies are in units of
  MHz.\\}\label{energylevels}
\end{figure}

Four well studied SNRs with strong 1720\,MHz masers (ranging from at
least 2.5 to 70 Jy) were selected (W\,44, W\,28, IC\,443 and
Sgr\,A\,East). Under project ID AP490, we have observed all
excited-state OH lines with excitation levels less than 500~K above
ground in the receiver bands\footnote{The currently ongoing upgrade to
  the Expanded VLA (EVLA) will also allow observations of the 6.0\,GHz
  lines, as well as the 13.4\,GHz lines later during the final stages
  of the upgrade. However, these transitions were not observable
  during this project. See \citet{fish07} for the 6.0\,GHz lines in
  these four SNRs.} of the VLA: the $\Lambda$-doubling triplet at
4.7\,GHz and the quadruplets at 7.8 and 8.2\,GHz (see Fig.\
\ref{energylevels}).  For Sgr\,A\,East we have also observed the
quadruplet around 23.8\,GHz (510~K above ground).  The 23.8\,GHz lines
were only observed in Sgr\,A\,East as this SNR lies in the region with
the highest known interstellar scattering, so observations in this
region would best benefit from a smaller $\lambda$. In addition, this
region is complex and dense, and might therefore provide the best
location to search for emission at 23.8\,GHz. The limited bandwidth
available at 23.8\,GHz ($\sim$36 km\,s$^{-1}$) may possibly exclude the
detections of masers with velocities much offset from the systemic LSR
velocity.

W\,28, W\,44 and Sgr\,A\,East were observed in CnB-configuration 2005
July 5, and IC\,443 was observed in the C-configuration 2005 September
5. Based on the positions of the strongest 1720\,MHz masers
\citep[$>2.5$\,Jy;][]{claussen97}, three regions were selected in
W\,28, two regions in W\,44 and Sgr\,A\,East respectively, and one
region in IC\,443. Given that the line widths of most detected
1720\,MHz masers are relatively narrow, we observed with a channel
separation of 12.207\,kHz ($\lesssim1$~km\,s$^{-1}$ at 4.7\,GHz) and a
bandwidth of 1.56\,MHz (which was changed to 48.828\,kHz separation at
3.1\,MHz bandwidth for the 23.8\,GHz observations only). Using 4~IF
(Intermediate Frequency) mode at the VLA, we could observe two
transitions within each band simultaneously in dual circular
polarization centered on the systemic local standard of rest (LSR)
velocity of the source. For the 4.7\,GHz triplet, we used the
redundant IF for the 4750\,MHz line, centered at another velocity (see
Table~1).

The data were calibrated using NRAO's Astronomical Image Processing
System (AIPS), and imaged with natural weighting using standard AIPS
procedures. If continuum emission existed in the field, this continuum
was subtracted in the UV-plane before imaging. In Table~1, we
summarize the results of the observations, including the pointing
positions. For each pointing position and frequency, we integrated on
source for 5 minutes. The typical final rms noise was $\leq10$~mJy per
channel, except for Sgr\,A\,East where it was $\leq15$~mJy due to
image fidelity issues in this complex region. The field of view at
4.7, 7.8, 8.2 and 23.8\,GHz was 9.6, 5.8, 5.5 and 1.9\arcmin\
respectively.

\subsection{Collisionally excited 1612 MHz OH masers}
\label{coll1612}
Models of collisional excitation that predict 1720~MHz maser emission
also predict 1612~MHz masers in regions of higher density
($n\sim10^7$\,cm$^{-3}$) or higher column density
\citep{pavlakis96}. Since 1612~MHz emission traces much denser
material than 1720\,MHz, 1612~MHz masers are not primarily expected in
the immediate vicinity of 1720~MHz masers. We therefore
cross-correlated the positions of known SNRs with the locations of
known 1612~MHz maser emission in the literature. The positions of the
SNRs were taken from the SNR catalogue of \citet{green06}, and the
1612~MHz masers were selected from blind surveys \citep{telintel91,
  sevenster97a, sevenster97b, sevenster01}. Since most 1612~MHz masers
with a double-peaked spectrum are associated with evolved stars
\citep[e.g.,][]{habing96}, we constrained our cross-correlation to
sources with single-peaked or irregular spectra, which resulted in a
final sample of 184 sources. No probable associations were found
within a minimum search radius of 10\arcmin, suggesting that the
higher column densities suitable to produce collisionally excited 1612
MHz SNR maser emission do not occur frequently in SNRs. A special case
may be Sgr\,A\,East in the Galactic center region where a much more
sensitive 1612\,MHz OH survey is available \citep{sjouwerman98}, and
which we will further discuss separately in Sect.\ \ref{gc}.

\section{Discussion}

\subsection{Other detections of excited-state OH lines}
No detection of either absorption or emission was made in any of the
lines searched for excited-state OH (but see \citet{fish07} for
possible 6030/6035\,MHz main line absorption in Sgr\,A\,East). This
may seem somewhat surprising, since modeling of the three lower
rotational states of OH in SFRs has demonstrated that 1720, 4765 and
6035\,MHz masers can occur under similar conditions in regions of
shocked gas \citep{gray91, gray92}. Observations of SFRs support the
coexistence of these lines. Indeed, a search for excited-state OH
masers at 4765 and 6035\,MHz in a sample of SFRs with 1720\,MHz masers
resulted in high detection rates \citep{macleod97}. About one-third of
the 1720\,MHz masers have associated 4765\,MHz masers, and as many as
about two-thirds display 6035\,MHz masers, indicating that the
excited-state OH masers may form under similar conditions to the
1720\,MHz masers. This is further supported by detailed mapping of the
SFR W\,3\,(OH), showing that a third of the 4765\,MHz spots are
spatially coincident with 1720\,MHz masers \citep{palmer03}. In
several SFRs, rotational lines as high as $\sim500$~K above ground (in
the seventh-lowest rotational level, $^2\Pi_{3/2}, J=9/2$) have been
detected in absorption and emission, including weak maser emission
\citep[e.g.\ in W\,3\,(OH) and Sgr\,B2][]{baudry81, gardner87,
  wilson90, baudry02, fish07}.

Most SFRs show main-line 1665/1667\,MHz masers, though there are
exceptions: \citet{niezurawska04} report on 4765\,MHz masers
associated with SFRs that display 1720\,MHz emission without main line
emission, suggesting that these regions could contain shocks with
similar properties to SNR shocks. Despite observations of coexisting
1720, 4765 and 6035\,MHz masers in SFRs, we have not been able to find
higher excitation lines in SNRs. In contrast to SNRs however, where
1720\,MHz masers are thought to be collisionally pumped, the main-line
1665/1667 and 6035\,MHz masers are probably radiatively excited. It
should be pointed out that the models by \citet{gray91,gray92}
predicting co-propagating 1720, 4765 and 6035\, MHz masers include a
strong far-infrared radiation field associated with the parent star,
suitable for SFRs. Hence, radiative pumping routes are likely to be
critical for producing simultaneous 1720, 4765\,MHz satellite line
emission and 6035\,MHz main line emission.

\subsection{Predicted OH maser transitions in SNRs} 

\begin{figure}[t]
\resizebox{9cm}{!}{\includegraphics{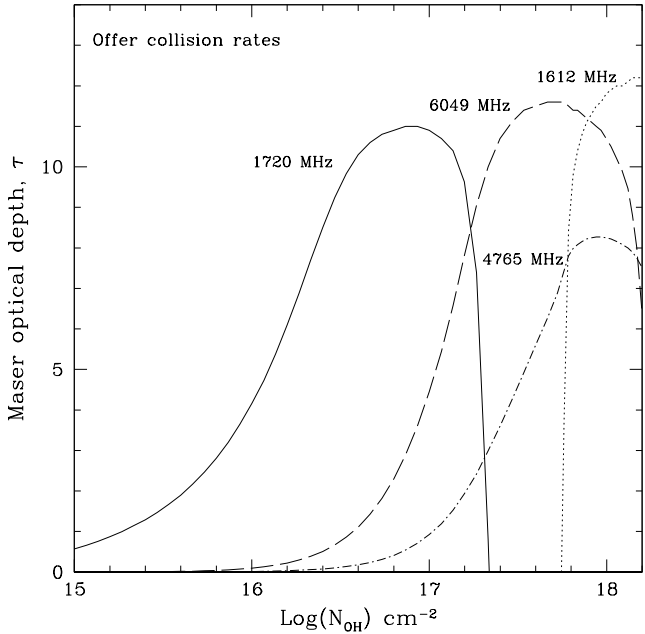}}
\caption{Expected maser optical depths in inverted transitions in a gas
  component typically producing 1720\,MHz masers ($n=10^5$\,cm$^{-3}$,
  $T=75$\,K, $f_{\rm OH}=10^{-5}$), using the \citet{offer94} collision
  rates.}\label{offer}
\end{figure}

Existing models predict a sequence of inversions as the number density
($n_{\mathrm{H}_2}$) or the column density of OH ($N_{\mathrm{OH}}$)
is increased \citep{pavlakis96,pavlakis00,lockett99,wardle07}. To
predict the maser optical depths in all 1.6, 4.7, 6.0, 7.8, 8.2 and
13.4\,GHz transitions, we have used MOLPOP\footnote{Available at
  http://www.pa.uky.edu/~moshe/molpop.zip}. The MOLPOP program solves
the molecular level population equations using the escape probability
method for a homogeneous slab. We use the \citet{offer94} collision
rate coefficients for the lowest 24 energy levels, with an ortho-para
ratio of 3:1. Thus, the 23.8\,GHz transitions are not included in this
model. Following the results by \citet{lockett99}, we start with
post-shock gas properties typically producing 1720\,MHz masers in
SNRs: molecular density $n_{H_2}\sim10^5$\,cm$^{-3}$, temperature
$T\sim75$\,K, OH fraction $f_{\rm OH}\sim10^{-5}$, and thermal
line-widths. Using these values, in Fig.\ \ref{offer} we plot the
maser optical depth as a function of OH column density, $N_{\rm
  OH}$. These results are consistent with those of
\citet{wardle07}. With the given parameters, the model does not
predict any significant maser optical depths in any of the 7.8, 8.2
and 13.4\,GHz lines, nor in the individual 1665, 1667, 4660, 4750,
6016, 6030 or 6035\,MHz lines. Note that the 6035\,MHz masers modeled
by \citet{gray91, gray92} thus must be due to IR pumping and therefore
will not be discussed further below. At low column densities,
1720\,MHz maser emission is produced. At progressively higher column
densities, first 6049\,MHz masers turn on, then 1720\,MHz masers turn
off, so that there is an overlap range in which the OH column density
is simultaneously low enough to produce a detectable 1720\,MHz maser
and high enough to produce 6049\,MHz maser emission as well. At still
higher column densities, the 4765 and 1612\,MHz masers turn on and the
6049\,MHz masers turn off. The transition between 1720\,MHz masing and
1612/4765\,MHz masing occurs when the 79\,$\mu$m transitions between
the $^2\Pi_{3/2}, J = 3/2$ states and the $^2\Pi_{1/2}, J = 1/2$
states (Fig.\ \ref{energylevels}) become optically thick
\citep{elitzur76}.  It is worth noting that in the low-temperature ($T
< 120$\,K) regime, the collisionally-excited masers occur only in
satellite-line transitions ($\Delta F = \pm 1$, see Fig.\
\ref{energylevels}). At higher temperatures, main lines ($\Delta F =
0$) can become inverted as well via collisions and local line overlap
\citep[e.g.,][]{pavlakis96, pavlakis00}.

Inversion of the 1720\,MHz maser using the Offer collision rates
depends on the hydrogen ortho-para ratio \citep{pavlakis96}. To test
the reliability of our results, we also calculated the level
populations using hard sphere cross sections and obtained results
similar to those found using the Offer rates with an ortho-para ratio
of 3.

The only OH transition observed to produce a detectable maser in SNRs
is the 1720\,MHz transition. A targeted search for 6049\,MHz masers
toward 36 SNRs failed to detect a single maser
\citep{mcdonnell07}. Recently, we also used the Effelsberg telescope
to search for the four 6\,GHz lines in W28, W44, IC\,443 and
Sgr\,A\,East, with no detections \citep{fish07}. As noted in Sect.\
2.1, the 1612\,MHz masers detected in the \citet{telintel91,
  sevenster97a,sevenster97b,sevenster01} blind surveys do not produce
any detections associated with SNRs (but see Sect.\
\ref{gc}). Finally, in this work, we fail to obtain positive
detections at 4765\,MHz or any other excited-state transition.

\subsection{Detectability of predicted masers}
One reason why no excited-state OH lines have been detected may be due
to sensitivity. The maser emission will be amplified according to
$T_{\rm m}\simeq T_{\rm bg}e^{\tau}$, where $T_{\rm m}$ is the
brightness temperature of the maser, $T_{\rm bg}$ is the brightness
temperature of the background continuum radiation, and positive $\tau$
is the maser optical depth. By comparing the brightness temperature of
a known 1720\,MHz maser and its background continuum, we can calculate
the maser optical depth and compare it to the values predicted in
Fig.\ \ref{offer}. If consistent, the results can be scaled to
estimate the predicted maser brightness temperatures in the excited
states.

As an example, we consider Sgr\,A\,East. We select the region where a
0.18~Jy maser occurs at +55~kms$^{-1}$ \citep{yusef-zadeh96}. At that
position, the background continuum at 1720\,MHz corresponds to a
brightness temperature of $T_{\rm bg}\sim 100$\,K, and the maser
brightness temperature is $T_{\rm m}\sim2.3\times10^5$\,K. Assuming
all continuum is in the background, this implies a lower limit to
optical depth of $\tau\sim8$, well in agreement with Fig.\ \ref{offer}
and other existing models \citep{lockett99, wardle07}. To estimate the
$T_{\rm m}$ for the transition at 4765\,MHz, the 1720\,MHz brightness
temperature is scaled using the spectral index $\alpha=-1$ measured in
Sgr\,A\,East by \citet{pedlar89}. The brightness temperature of a
synchrotron emitter changes as $T_{\rm b}\propto\lambda^{2-\alpha}$
\citep{rybicki79}, yielding a 4765 $T_{\rm bg}\sim5$\,K. An optimistic
reading of Fig.\ \ref{offer} in the region where 1720\,MHz emission
still occurs will give a $\tau\sim2$, resulting in $T_{\rm m}\sim
40$\,K. Indeed, this is below the 5-$\sigma$ detection limit in our
VLA observations, which is 50\,mJy or 115\,K. Thus, it might be
difficult to detect 4765\,MHz masers co-propagating with 1720\,MHz
masers without deeper integrations than those presented in Sect.\ 2.

A similar estimate for the 6049\,MHz line, assuming a $\tau\sim6$ and
$T_{\rm bg}\sim2.5$\,K yields $T_{\rm m}\sim1\times10^3$\,K, which
should be easier to detect. With the upgrade of the VLA to EVLA, the
6.0\,GHz lines will be observable and are thus warranted a deeper
search. However, we note that to date, searches for 6049\,MHz have
proved negative. This indicates that the 1720\,MHz masers occur only
in regions of low column densities, insufficient for the formation of
4765 and 6049\,MHz masers.

\subsection{Column density effects} 

The most straightforward explanation of the absence of detectable
excited-state OH maser emission in SNRs is that highest column density
peaks produce the 1720\,MHz masers but are not high enough to generate
maser emission in any other transition. We note that the search
presented here is biased toward regions with existing 1720\,MHz
masers, and could thus be biased against regions with the higher
column densities needed for the excited-state OH (Fig.\
\ref{offer}). Estimates of the OH column densities are available via
thermal absorption in the ground state transitions of OH, when
sufficient background continuum emission is present. Absorption of
1667\,MHz OH in W\,28 implies $N_{\rm OH}\simeq
2\times10^{16}$\,cm$^{-2}$, for an adopted excitation temperature
$T_{\rm ex}$ of 10\,K \citep{yusef-zadeh03a}, indicating low column
densities with little possible inversion of the excited-state OH
(Fig.\ \ref{offer}). Similar observations of IC\,443 show OH
absorption in molecular clumps with $N_{\rm OH}$ ranging between
$0.7\times10^{16}-1.4\times10^{17}$\,cm$^{-2}$
\citep{hewitt06}. Slightly higher column densities of the order of
$2-3\times10^{17}$\,cm$^{-2}$, assuming $T_{\rm ex}\simeq10$\,K, are
derived for Sgr\,A\,East (Sjouwerman, priv.\ comm). With measured OH
column densities of the order of $10^{16}-10^{17}$\,cm$^{-2}$, models
based on the \citet{offer94} collision rates allow the possibility
that there might exist a few regions in SNRs with sufficient column
densities for the 6049\,MHz line to be inverted (Fig.\
\ref{offer}). Our non-detections are, however, consistent with regions
of lower column densities, where 1720\,MHz masers are strong, and
little or no inversion has occurred in the higher transitions.

\subsection{Temperature and density}

An alternative explanation to the absence of excited-state OH masers
could be that the density or temperature is too high. Models of
1720\,MHz emission imply that the optimal post-shock density in which
1720\,MHz masers occur is of the order $n_{\rm
  H_2}\sim10^5$\,cm$^{-3}$, with temperatures $T\sim50-125$\,K
\citep{lockett99, wardle99}. Millimeter observations of multiple
molecular lines in IC\,443 suggest that there may be regions of much
higher temperature and density in SNRs. In particular, in IC\,443, the
1720\,MHz masers occur within the IC\,443 cloud complex G
\citep{denoyer79}. From the millimeter data, the medium in this clump
appears to fit a two-component model with both a cool, low-density
($T\sim80$\,K, $n\sim10^5$\,cm$^{-3}$) component, and a warmer, higher
density ($T\sim200$K, $n\sim3\times10^6$\,cm$^{-3}$) component
\citep{vandishoeck93}. Similar numbers were derived by
\citet{turner92}. At densities of the order of $10^6$\,cm$^{-3}$, but
with lower temperature ($T=75$\,K), previous modeling has already
shown that the 1720\,MHz transition may still be weakly inverted
\citep{lockett99}. 

\begin{figure}[t]
\resizebox{9cm}{!}{\includegraphics{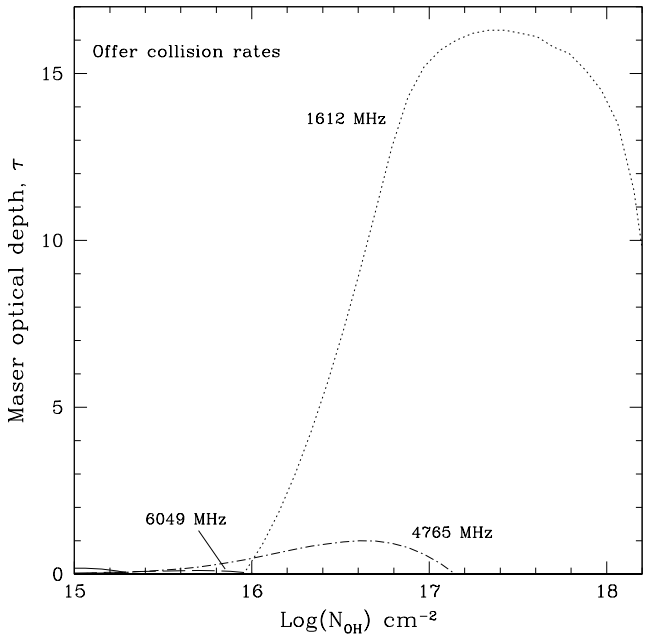}}
\caption{Expected maser optical depths in the inverted transitions in
  a high density ($n=5\times 10^6$\,cm$^{-3}$, $T=200$\,K) gas
  component. At these high temperatures and densities, the 1720~MHz
  maser is quenched and the 1612~MHz line dominates.\\}\label{mpdense}
\end{figure}

Here we calculate the expected maser optical depths in a warmer,
higher density region, again using MOLPOP assuming thermal
line-widths, $T\simeq200$\,K, and $n\sim5\times10^6$cm$^{-3}$, we
achieve the results shown in Figure \ref{mpdense}. As expected, the
1720\,MHz line is quenched at this higher density. The 4765\,MHz
transition is only modestly inverted with maser optical depths of the
order of 0.5--1 at column densities $N_{\rm
  OH}\sim10^{16}-10^{17}$\,cm$^{-3}$. Even weaker inversion is seen in
the 6049\,MHz line. We therefore conclude that if the molecular medium
is commonly composed of two components like those in IC\,443, the
1720\,MHz masers must be associated with the cooler, lower-density
component. A warmer component will not be able to produce
excited-state OH masers, and as a consequence, we conclude that it is
not the density that primarily constrains the formation of
excited-state OH.

We note that at higher densities, the model predicts a strong maser
line in the 1612\,MHz transition (Fig.\ \ref{mpdense}). However,
1612\,MHz masers in general are not correlated with SNRs (Sect.\ 2.2),
which could indicate that warm, dense regions like the one in
IC\,443\,G modeled by \citet{vandishoeck93} are rare in SNRs.

\subsection{A possible exception: Sgr\,A\,East}\label{gc}

Sgr A East is a very complicated region and has been studied in detail
in the past. In particular, it has been surveyed for 1612\,MHz masers
to find OH/IR stars \citep[][and references
therein]{sjouwerman98}. However, 1612\,MHz emission can also originate
from the interaction of a SNR and its surrounding ISM (Sect.\
\ref{coll1612}; Fig.\ \ref{mpdense}). We therefore turned to the list
of double peaked 1612\,MHz masers found by \citet{sjouwerman98}, where
some maser sources are attributed to the $+$50\,km\,s$^{-1}$ molecular
cloud (MC). The individual MC spectra shown may not represent the true
emission (and absorption) in the cloud, nor be complete for emission
of the cloud (e.g., see the ``maxmap'' in their Fig.~8) as the data
reduction and search were focused on finding double peaked OH/IR
stars. However, their MC list does partly represent compact 1612\,MHz
emission, either thermal or low-gain masing not attributed to stars,
and therefore is relevant in this respect.

In Fig.\ \ref{sgraeast} we plot the locations of the pure MC 1612\,MHz
emission (circles) from \citet{sjouwerman98} on top of the 1720\,MHz
SNR masers (crosses) and 1.7\,GHz radio continuum (grey scale) taken from
\citet{pihlstrom06}. The following observations can be made:

\begin{figure}[t]
\resizebox{9cm}{!}{\includegraphics{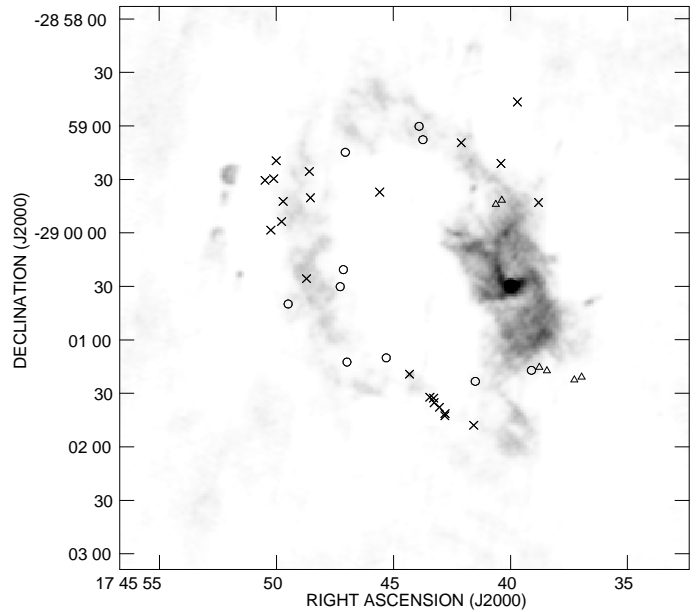}}
\caption{Sgr\,A\,East 1720\,MHz masers (crosses) superimposed on
  1.7\,GHz continuum image (grey scale) taken from
  \citet{pihlstrom06}. The triangles are high-velocity 1720\,MHz
  masers related to the circumnuclear disk and not to the SNR/MC
  interaction. The circle denotes positions of 1612\,MHz emission in
  this region.\\}\label{sgraeast}
\end{figure}

\begin{itemize}
\item[-] All regions of 1612\,MHz emission have one or more peaks in
  the V$_\mathrm{LSR}$ range of 30 to 70 km\,s$^{-1}$, i.e., closely
  following the V$_\mathrm{LSR}$ gradient over, and range of the
  $+$50\,km\,s$^{-1}$ molecular cloud.

\item[-] The 1612\,MHz emission avoids regions of 1720\,MHz emission
  (the closest possible association is about 0.5\,pc in projection),
  which is to be expected as a reflection of column density
  differences (Figs.\ \ref{offer} \& \ref{mpdense}).

\item[-] The 1612\,MHz emission is co-located with the radio
  continuum outlining the SNR, but not on top of the regions of the 
  strongest radio continuum emission.

\item[-] No 1612\,MHz emission is seen near the compact \ion{H}{2}
  regions in the east and south: i.e., the 1612\,MHz emission is
  probably not related to SFRs.

\end{itemize}

We note that no 4765\,MHz emission (this work) nor 6049\,MHz emission
\citep{fish07} is detected toward Sgr\,A\,East, so there does not
exist a continuous density gradient where the 1720, 6049, 4765 and
then 1612\,MHz transitions, respectively, are purely collisionally
pumped. Based on these data, we are not able to determine whether the
1612\,MHz emission is purely thermal emission from the cloud. It could
also be pumped by the interstellar radiation field and perhaps
amplifying the background radio continuum, or indeed collisionally
pumped stimulated emission resulting from the reverse shock in the
SNR. To address these questions in the Galactic center (GC), a more
specific observational setup will be required to properly account for
details like, for example, missing zero-spacing flux. As the GC is a
special case, it is beyond the scope of this paper. Pending a more
detailed analysis of the GC region we therefore assume that there
generally is no collisionally excited 1612\,MHz emission due to SNR/MC
interactions, with a possible exception in the GC.

\section{Concluding remarks}
For OH column densities $N_{\rm OH}\sim 10^{16}-10^{18}$ cm$^{-2}$,
models of collisionally-pumped excited-state OH in SNRs predict
inversions in the 1720, 4765 and 6049\,MHz lines, while no significant
maser optical depths will be produced in other transitions of the 4.7,
7.8, 8.2 and 23.8\,GHz lines.  We have presented the results from a
search of all these excited-state OH transitions in four, well-known
SNRs, with no detections. These VLA observations could not tune to the
6049\,MHz line, which is also predicted to have large optical depths
under conditions similar to those of 1720\,MHz masers. However, a few
recent searches for the 6049\,MHz line in SNRs have yielded no
detections \citep{fish07, mcdonnell07}. In the future, the EVLA
upgrade will allow for a targeted, deeper search for 6049\,MHz in
these SNRs.

Our non-detections are consistent with regions of lower column
densities ($N_{\rm OH}\leq5\times10^{16}$cm$^{-2}$), where 1720\,MHz
masers are strong, and little inversion occurs in the higher
transitions. Based on VLA and single-dish observations by
\citet{yusef-zadeh95} \& \citet{hewitt07}, such a post-shock medium
may have a large filling factor. \citet{hewitt07} found that a large
part of the 1720\,MHz maser flux is undetected using VLA baselines,
indicating a widespread distribution of weaker 1720\,MHz
emission. This likely reflects large post-shock regions of low column
density, or alternatively, regions of different temperature and
density from what is normally expected for 1720\,MHz maser
production. In turn, such a gas component will provide little chance
of excited-state OH maser emission. If this is the case, excited-state
OH will not be detected either coincident with or offset from detected
1720\,MHz masers in SNRs. The non-detections imply a low OH column
density in SNRs, in agreement with the rather low estimates of the
$N_{\rm OH}\simeq 10^{16}$\,cm$^{-2}$ expected to be produced in
C-type shocks \citep{wardle99, lockett99}.

From these results we draw the conclusion that in the regions where
1720\,MHz masers are found, the OH column densities are insufficient
for excited-state OH masers to exist. This is supported by the absence
of 1612\,MHz masers in SNRs, which require 1--2 orders of magnitude
higher column densities than the 1720\,MHz masers. However, the upper
limit to the column density must be even stricter to avoid producing
6049 and 4765\,MHz masers.


\acknowledgments 

The National Radio Astronomy Observatory is a facility of the
National Science Foundation operated under cooperative agreement by
Associated Universities, Inc.

{\it Facilities: \facility{VLA}}

\end{document}